# BMC Genomics

Methodology article

# Conversion of cDNA differential display results (DDRT-PCR) into quantitative transcription profiles

Balakrishnan Venkatesh, Ursula Hettwer, Birger Koopmann and Petr Karlovsky*

Address: Institute of Plant Pathology and Plant Protection, Goettingen University, Grisebachstrasse 6, D-37077 Goettingen, Germany

Email: Balakrishnan Venkatesh - venkey4@rediffmail.com; Ursula Hettwer - uhettwe@gwdg.de; Birger Koopmann - bkoopma@gwdg.de; Petr Karlovsky* - karlovsky@web.de

* Corresponding author

## Abstract

**Background:** Gene expression studies on non-model organisms require open-end strategies for transcription profiling. Gel-based analysis of cDNA fragments allows to detect alterations in gene expression for genes which have neither been sequenced yet nor are available in cDNA libraries. Commonly used protocols for gel-based transcript profiling are cDNA differential display (DDRT-PCR) and cDNA-AFLP. Both methods have been used merely as qualitative gene discovery tools so far.

**Results:** We developed procedures for the conversion of cDNA Differential Display data into quantitative transcription profiles. Amplified cDNA fragments are separated on a DNA sequencer and detector signals are converted into virtual gel images suitable for semi-automatic analysis. Data processing consists of four steps: (i) cDNA bands in lanes corresponding to samples treated with the same primer combination are matched in order to identify fragments originating from the same transcript, (ii) intensity of bands is determined by densitometry, (iii) densitometric values are normalized, and (iv) intensity ratio is calculated for each pair of corresponding bands. Transcription profiles are represented by sets of intensity ratios (control vs. treatment) for cDNA fragments defined by primer combination and DNA mobility. We demonstrated the procedure by analyzing DDRT-PCR data on the effect of secondary metabolites of oilseed rape *Brassica napus* on the transcriptome of the pathogenic fungus *Leptosphaeria maculans*.

**Conclusion:** We developed a data processing procedure for the quantitative analysis of amplified cDNA fragments separated by electrophoresis. The system utilizes common software and provides an open-end alternative to DNA microarray analysis of the transcriptome. It is expected to work equally well with DDRT-PCR and cDNA-AFLP data and be useful particularly in reseach on organisms for which microarray analysis is not available or economical.

## Background

Transcriptome analysis is a common way of discovering differences in gene expression because regulation of gene activity occurs primarily on transcription level. Numerous low-cost, simple methods are available for gene discovery projects, providing a limited set of transcripts more or less randomly selected from a pool of genes expressed differently between two samples (for example treated sample/control or diseased/healthy tissue). These methods include differential hybridization, subtractive hybridization, EST sequencing and other gene discovery methods, which do not provide quantitative data nor do they aim at a complete coverage of the transcriptome.

DNA microarrays became popular as a tool for genome-wide transcription analysis. Microarray analysis of gene expression is only feasible when extensive sequence information and/or cDNA libraries are available. Therefore, DNA microarrays do not satisfy a growing demand for "open-end" transcriptome analysis. Proprietary high-end transcript profiling systems such as Massive Parallel Signature Sequencing [1] and GeneCalling [2] are excellent tools for genome-wide quantitative transcription analysis on any organism, but they are available to only a small fraction of researchers. Serial Analysis of Gene Expression [3] is a free, open-end, quantitative transcription profiling system. The drawback of SAGE is that the size of the sequencing effort required to generate representative data sets limits the practical applicability of the method, particularly when transcripts expressed at low levels need to be covered.

mRNA differential display (DDRT-PCR) [4], cDNA-AFLP [5] and different variants of these two methods [6] are gel-based transcript profiling systems based on electrophoretic fingerprinting of amplified cDNA fragments. Both DDRT-PCR and cDNA-AFLP are open-end methods requiring only standard instrumentation and incurring low costs. They have been used in hundreds of published studies both on model and non-model organisms so far. The reproducibility of cDNA-AFLP patterns was reported to be superior to DDRT-PCR, but its drawback is a high fraction of cDNA molecules escaping detection because of the lack of suitable restriction sites [7]. Because of this limitation and a higher technical demand for cDNA-ALFP analysis as compared to DDRT-PCR, the latter method has largely dominated gel-based transcription profiling [6].

The fundamental goal of transcriptomics is to generate and compare snapshots of mRNA populations. In order to be suitable for comparisons with new data sets, these transcription profiles have to consist of quantitative values stored in a digital form and allowing for normalization procedures compensating for differences in the efficiency of mRNA extraction, cDNA synthesis and detection. Neither DDRT-PCR nor cDNA-AFLP in their established forms fulfill these requirements. So far both methods have been used for gene discovery based on subjective band selection [8], though different steps towards automatic large-scale transcription profiling have been undertaken, including the use of DNA sequencers for cDNA fragment analysis [9-12], linking expression data to sequences [13] and quantitative evaluation of autoradiographs of cDNA-AFLP gels [7].

Here we describe procedures for the conversion of fluorescent differential display results into quantitative transcription profiles, using a DNA sequencer for cDNA fragment separation and DNA fingerprinting software for band matching, data normalization and densitometry. Transcription profiles are represented by sets of intensity ratios calculated for band pairs (treated sample vs. control) defined by the primer combination used and DNA fragment mobility. We demonstrate the method by analyzing the effect of secondary metabolites of oilseed rape *Brassica napus* on the transcriptome of the rape pathogen *Leptosphaeria maculans*. To our knowledge this is the first report presenting a quantitative evaluation of a mRNA differential display experiment.

# Results
## *Frequency and size distribution of cDNA amplicons*
DDRT-PCR partitions transcripts into non-overlapping sets by synthetising cDNA and amplifying their terminal parts with primer pairs consisting of an anchored poly(dT)-primer and a random primer. We used three anchored primers and 76 random primers (see Arbitrary_primers.xls), resulting in 228 primer combinations. Amplification products were separated on a DNA sequencer and detector signals were converted into virtual images (Fig. 1). A total of about 10,000 distinct fragments of *Leptosphaeria maculans* cDNA were detected. 5,422 bands exceeded the thresholds set for the size (50 bp) and intensity (0.5% elevation as compared to the surrounding). The average number of detected bands produced by one primer pair were 73, the minimum were 0 and the maximum were 263 bands per primer pair. 65% of arbitrary primers produced 40 – 180 bands each (Fig. 2).

Frequencies of bands produced with different anchored primers are compared in Fig. 3. All three anchored primers produced a similar number of bands, indicating that there is no preference for the last non-T nucleotide in the poly-adenylation site of mRNA in *Leptosphaeria maculans*.

The size distribution of amplified fragments (Fig. 4) shows that short fragments were amplified preferably, which is typical of methods based on randomly primed PCR. Sequencing very short DDRT-PCR fragments is likely to detect 3'-noncoding sequences, which are not informative. Fragments exceeding the length of 200 nt, which would preferable be chosen for cloning and sequencing, made up a half of all evaluated fragments.

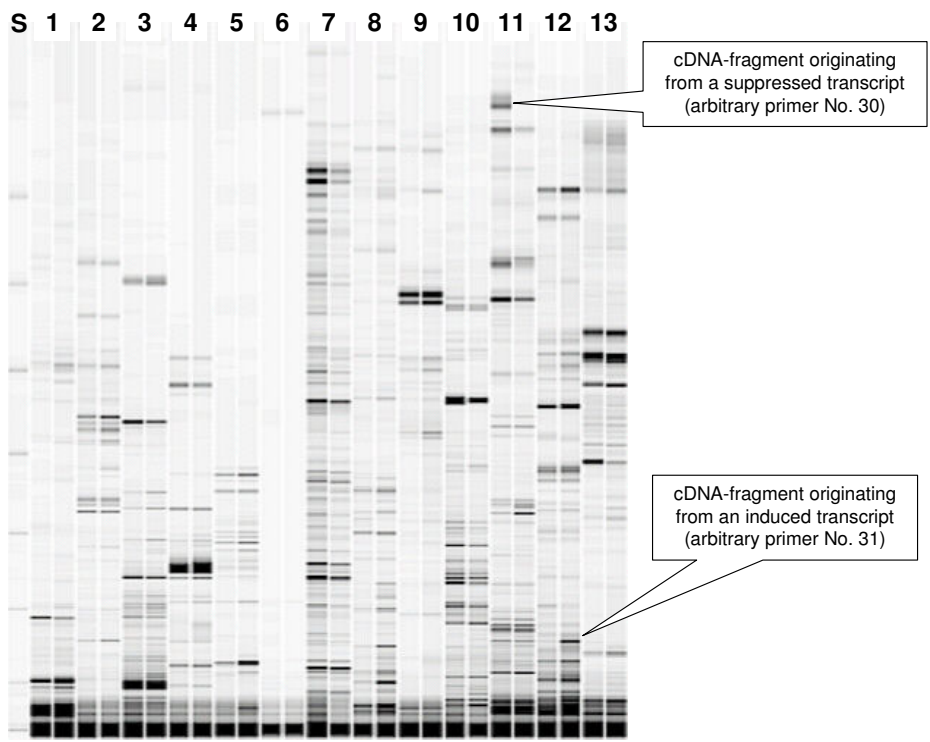

**Figure 1**
**Virtual gel image of DDRT-PCR products**. *Leptosphaeria maculans* cultures were treated with secondary metabolites of *Brassica napus*, RNA was extracted and DDRT-PCR reactions performed as described in Materials and Methods section. Amplicons labeled with Cy5 fluorophore were separated on a DNA sequencer and detector signals were converted into a virtual gel image. Thirteen primer combinations were used for the amplification of paired samples consisting of a control (left) and a treated culture (right) each. A size standard was loaded into the first lane.

*Coverage and reproducibility*
To accurately estimate the coverage of the DDRT-PCR experiment, one needs to know the number of genes expressed under culture conditions used. Because the genome of *L. maculans* has not been sequenced yet, we assume by analogy with other filamentous fungi that it harbours about 10 000 genes. Taking into consideration that many functions are not expressed in mycelial cultures used as the source of mRNA (e.g., activities related to mating, pathogenesis, sporogenesis and spore germination), we assume that the transcriptome analyzed in this work consisted of 4000 to 6000 mRNA molecules. DDRT-PCR generated 10000 cDNA bands, 5422 of which were selected for evaluation. Assuming Poisson distribution, this corresponds to a 59 – 75% coverage of the transcrip-tome. The coverage of published DD and cDNA-AFLP experiments varies from less than 20% to 73%.

It has been reported that DDRT-PCR results are afflicted with a relatively high proportion of false positives [6], which originate mainly from ribosomal RNA. The reproducibility of quantitative changes in transcription levels determined by DDRT-PCR has not been reported so far. The fact that induction factors for the vast majority of data points calculated from normalized band intensities were close to 1.0 (Fig. 5) indicates that the reproducibility of DDRT-PCR patterns was fairly good in this experiment, though we did not know which part of the differences is attributable to induction/suppression and which to random variance. To determine the reproducibility of the

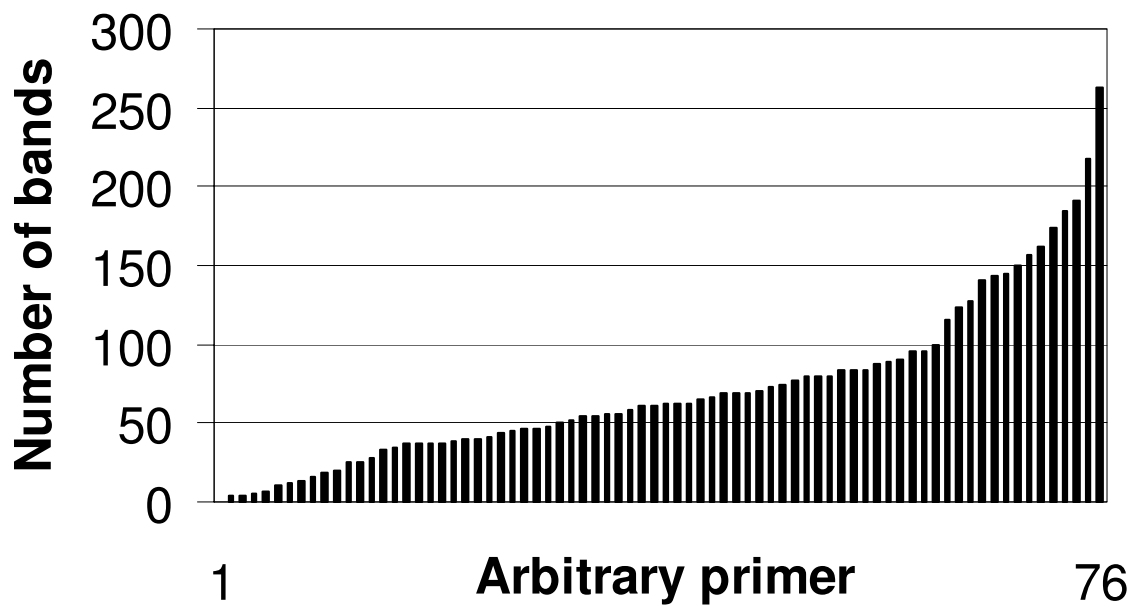

**Figure 2**
**Number of bands produced by 76 arbitrary primers**. Arbitrary primers are sorted by the total number of bands quantitative scoring, we repeated a set of DDRT-PCR reactions on RNA extracted from a second untreated (control) culture. Both cultures were inoculated with the same inoculum and grown in the same incubator to keep the differences in culture conditions at minimum. RNA extraction, cDNA synthesis and DDRT-PCR were performed in parallel, too, and care was taken to limit the variance introduced by handling. Band intensities were normalized and the ratios of intensities for matching bands calculated. The results are summarized in Tab. 2. These data confirmed that reproducibility is an important issue in DDRT-PCR, as is the case with microarrays and other profiling methods. The variance in our experiment was comparable to microarrays [cp. Table 3 in Ref. [14]].

***Changes in transcript levels after treatment of fungal cultures with phytoalexins***
In order to detect specific effects of *Brassica napus* phytoalexins on *Leptosphaeria maculans* transcriptome and exclude general stress responses, we chose very mild treatment conditions. Firstly, only a small fraction of *Brassica napus* metabolites was used for the treatment. In the purification protocol for these metabolites from plant material we combined two liquid-liquid extractions (from phosphate buffer at neutral pH into ethyl acetate and subsequently from ethyl acetate into hexane). Secondly, the concentration of metabolites in fungal culture medium was adjusted to be at most equal to their concentration in *Brassica napus* stem. Furthermore, a short treatment time was chosen to preclude secondary effects.

DNA yield for each evaluated cDNA fragment was determined as a densitometric value of the corresponding band, normalized to account for differences in loading volumes and used to calculate ratios between treated sample and control, designated as induction factor (see Materials and Methods for details). Fig. 5 summarizes induction factor values for all 5,422 evaluated bands. The

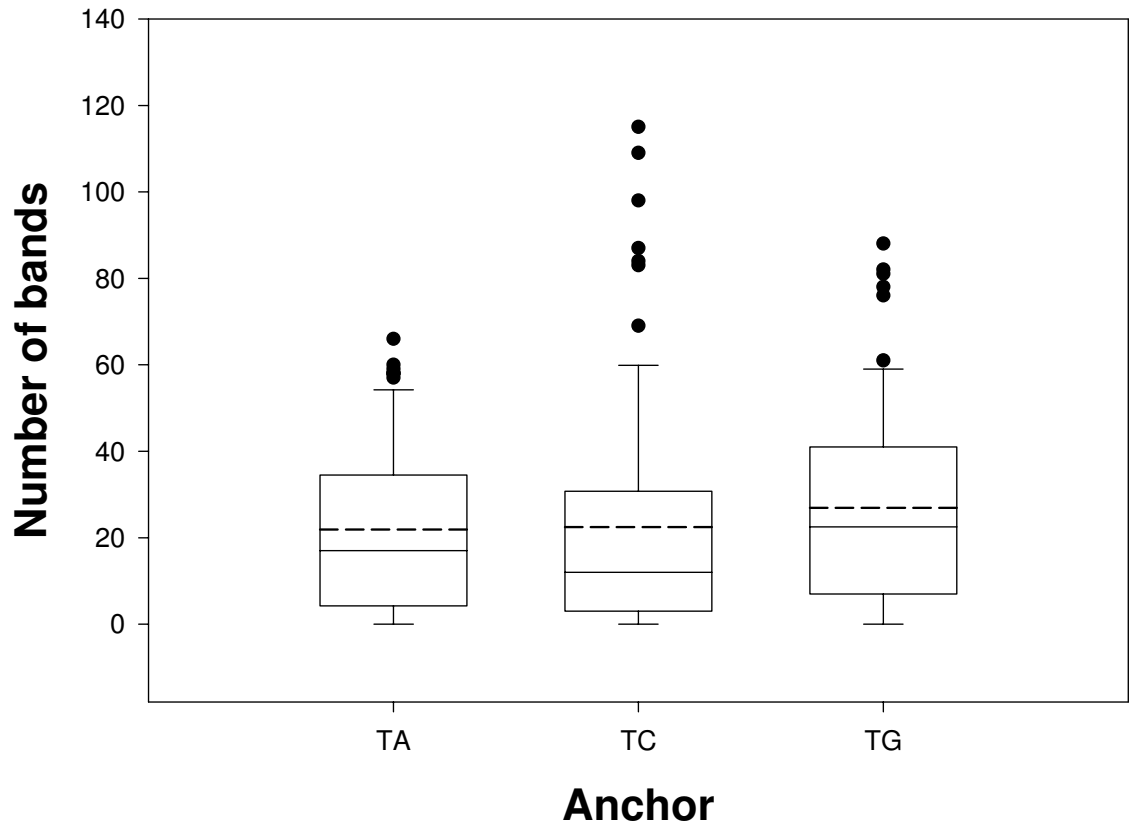

**Figure 3**
**Number of bands produced with anchored primers (T)$_{17}$A, (T)$_{17}$C, and (T)$_{17}$G**. Box-and-Whisker-plots: dashed line = mean, solid line = median, dots = outliers intensity of the majority of bands has changed to a lesser degree than three-times after the treatment. The induction factor was used to divide evaluated bands into five classes. As shown in Fig. 6, intensities of 174 bands (0.03 %) increased 3-times to 9-times upon treatment and 16 bands (0.003 %) were induced more than 9-times, the strongest induction observed being 19.5-times. In a gene discovery project these bands would represent candidates for cloning.

In addition to bands appearing in both treated sample and control, a few bands were detected only in the control or only in the treated sample, representing transcripts which were either not present in detectable amounts in the control or fully suppressed by the treatment. These bands cannot be characterized by an induction factor, which amounts to infinity for induced bands without a counterpart in the control and zero for bands fully suppressed in the treated sample. In order to further assess the significance of these bands, we compared their intensities with the intensities of all evaluated bands in the experiment (Table 1). The strongest band (present only in the treated sample) amounted to 2451% of the mean peak intensity calculated for all bands in the experiment. The strongest suppressed band (present only in the control) amounted to 951% of the mean peak intensity.

## Discussion
The main drawback of cDNA differential display as compared to DNA microarrays and industrial gel-based transcription profiling systems is a subjective evaluation of gels and the fact that data are not available in a quantitative form suitable for database storage and numerical manipulations. Though different steps towards quantitative evaluation of DDRT-PCR gels have been undertaken, a practicable system for DDRT-PCR-based transcriptomics based on common equipment has not been described yet.

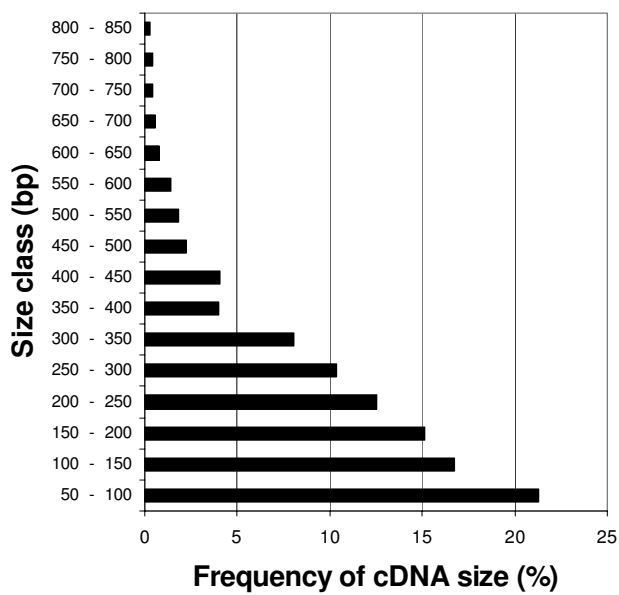

**Figure 4**
**Size distribution of amplified cDNA fragments**. Bands corresponding to fragments shorter than 50 bp were discarded. The size of detected bands ranged from 50 to 847 bp.

The system for quantitative evaluation of gel-based transcript profiles developed in this work relies on widely available equipment (DNA sequencer) and software (DNA fingerprinting analysis and spreadsheets). A crucial part of data processing is normalization, which compensates for differences in pipetting and in the efficiency of different steps from RNA extraction up to loading samples on the sequencer. A central issue in normalization is the choice of bands originated from transcripts unchanged by the treatment. Induced and suppressed bands have to be excluded from the calculation of a normalization factor, yet accurate distinction between induced/suppressed and constitutive bands is only possible with normalized data. We solved this problem by assuming that the majority of bands in each lane are not affected by the treatment and calculating a normalization factor from the middle quartiles of bands sorted by their uncorrected induction factor UIF (see Materials and Methods for details.) The efficiency of this approach was proven by the fact that the vast majority of induction factors calculated from normalized data were close to 1.0 (see Fig. 5). This result also demon-strated an excellent reproducibility of differential display patterns recorded on a DNA sequencer.

cDNA fragment patterns were originally recorded by autoradiography of radioactively labeled DNA separated in polyacrylamide gels [4] and radioactive labeling is still being used [7]. However, irregular background and typical artifacts of autoradiographs hampers automatic gel evulation. Bauer et al. [20] attempted to replace autoradiography by labeling cDNA fragments with fluorophores and separating them on a DNA sequencer. They used different fluorophores for controls and treated samples; because of the differences in fluorescence yield among fluorophores, quantification of the amount of DNA in cDNA bands was not possible. Therefore, the authors depicted their protocol as merely "a qualitative method allowing to identify genes which are completely switched on or off".

In further development of fluorescent DDRT-PCR, labeling random primers [12] was replaced by labeling anchored primers and capillary DNA sequencers were used for cDNA fragment detection [9,10,23]. Although the use of Genotyper Genetic Analysis software (Perking-Elmer ABI) facilitated semi-automatic recognition of unique bands (e.g., bands present in the treated sample but missing in the control), quantitative evaluation of data has not been demonstrated.

Aittokallio et al. [21,24] developed mathematical algorithms for the evaluation of DDRT-PCR patterns recorded on an ALFExpress II sequencer (the same system used in this work) as an alternative to transcript profiling by microarrays. Their algorithms were written from scratch, comprising functions for fitting and smoothing densitometry traces, automatic band finding, band matching (called alignment) and eventually identifying differently expressed bands. Their work provides a solid basis for the development of software for automatic, quantitative processing of DDRT-PCR results, but it is not mature yet. The software is not freely available and it is not clear from the publication whether it is user-friendly enough to be used outside the laboratory of its creators. Finally, certain aspects of the signal processing procedure needs scrutiny. For example, the difference in mobility between corresponding bands (treatment/control) affects the value of Matushita distance (equation 5 in [21]) which is used for the computation of the dissimilarity score (equation 8 in [21]), though these differences reflect merely physical conditions during gel electrophoresis and are unrelated to transcription. In our procedure we completely separated band matching from the comparison of band intensities. Another potential problem with signal processing in [21] is the compensation for differences in loading volumes between lanes. Instead of normalizing lane intensities, the authors attempted to achieve a "similar background" by dividing the "distance value" for each peak pair by the average distance of all other peak pairs in the lane. This procedure is likely to produce erroneous results when two or more bands in a lane originate from transcripts of differentially expressed genes. Furthermore, it will introduce a bias into distance values of constitutively expressed (unchanged) transcripts in lanes which contain strongly induced or suppressed bands. We avoided this problem by excluding 25% bands with the largest and 25% bands with the smallest induction factor from the computation of the normalization factor (see Materials and Methods for details).

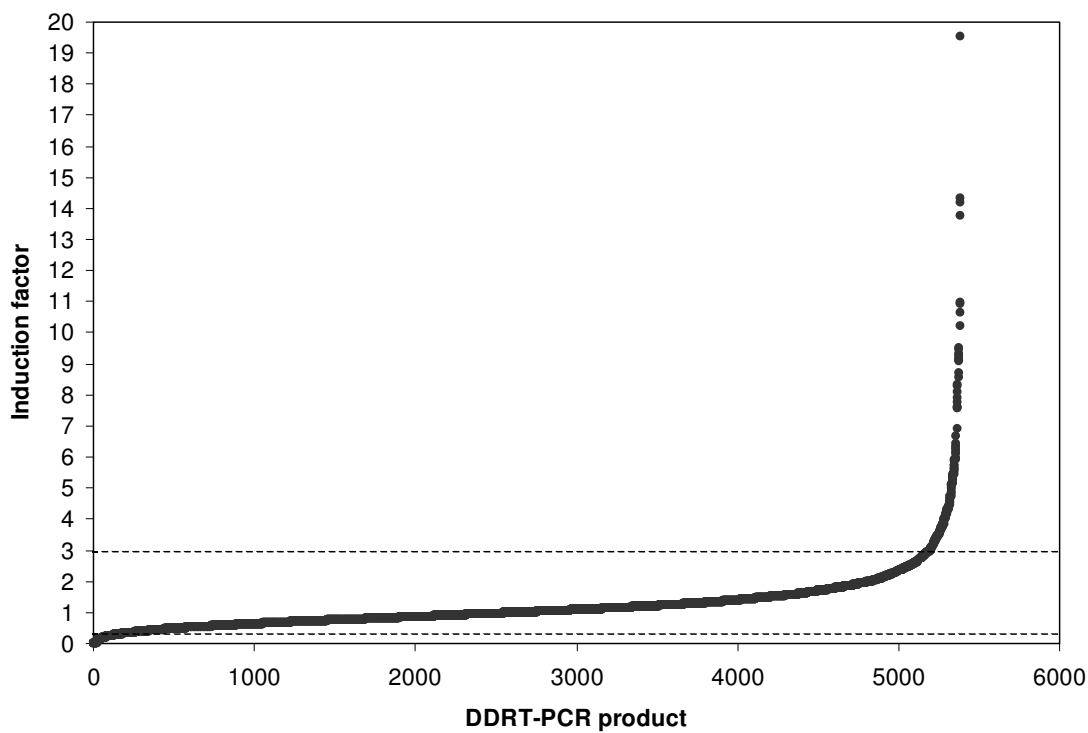

**Figure 5**
**Changes in intensity of DDRT-PCR bands from mRNA of *Leptosphaeria maculans* after treatment with *Brassica napus* extract**. Bands which passed thresholds of relative intensity and elevation from background (see Materials and Methods), sorted by the value of their induction factor (the ratio of normalized intensities of treated sample vs. control), are listed on the abscissa. Vertical lines drawn through the curve correspond to induction factor values of 3.00 and 0.33 (induction or repression by a factor of 3, respectively).

Gellatly et al. [22] recently studied the effect of methyl-benzyldithiocarbamate (analog of phytoalexin brassinin) on *Leptosphaeria maculans*. They used DDRT-PCR and observed on the average 65 bands for each primer combination, which is similar to our results. Because the majority of gel-excised bands could not be amplified, the authors later abandonded DDRT-PCR and swiched to cDNA-AFLP. We successfully amplified and cloned bands selected from DDRT-PCR gels for *Leptosphaeria maculans* by visualizing Cy5-labeled DNA on a fluorescence scanner (to be published). There are two differences in the treatment conditions used by Gellatly at al. [22] as compared to our work. The first difference concerns the composition of the inducing agent: a defined brassinin analogue was used by Gellatly [22] while we extracted metabolites from

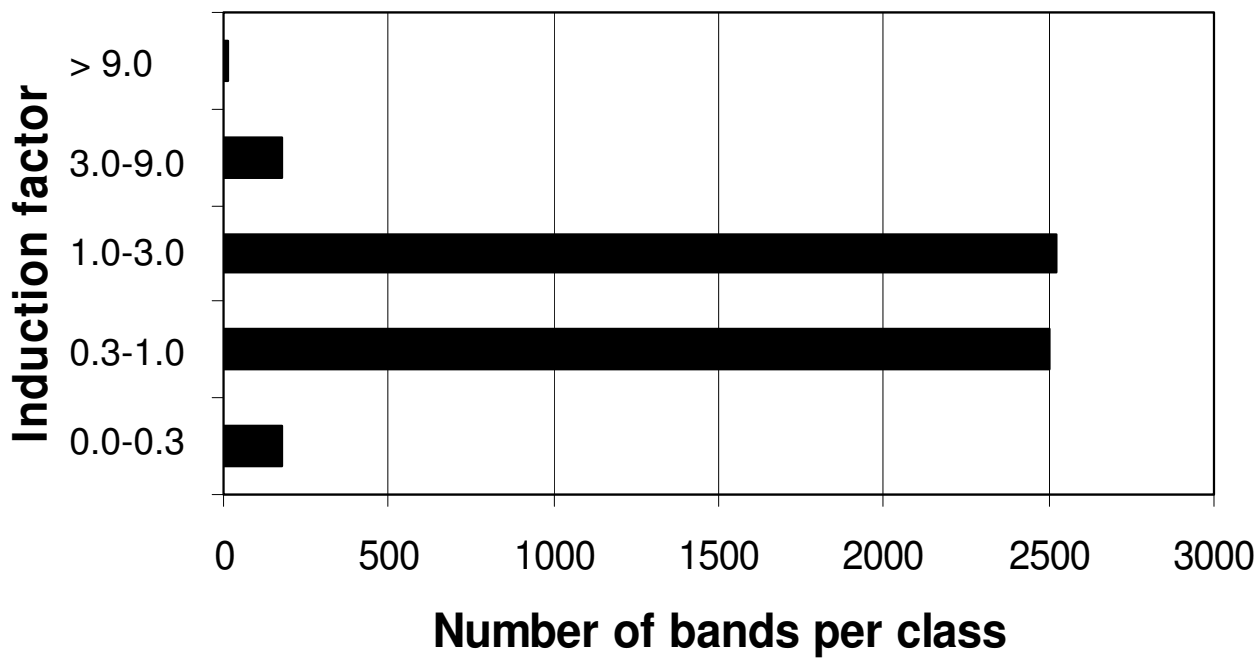

**Figure 6**
**Classification of cDNA bands in induction classes**. Evaluated cDNA bands were assigned to five classes according to the ratios of intensities sample/control.

**Table 1: Solitary cDNA fragments**

Bands detected in treated sample but missing in the control

| Relative intensity* | Number of bands |
|---|---|
| 0 – 30% | 3 |
| 30% – 300% | 19 |
| > 300% | 10 |

Bands detected in the control but missing in treated sample

| Relative intensity* | Number of bands |
|---|---|
| 0 – 30% | 3 |
| 30% – 300% | 13 |
| > 300% | 6 |

*as compared to the average of densitometric values of all bands in the experiment DDRT-PCR bands in *L. maculans* cultures treated with *Brassica napus* secondary metabolites vs. untreated control were matched and their intensities compared (Fig. 5 and 6). For bands with no counterpart, the relative intensity as compared to the average intensity of all bands in the experiment was determined.

*Brassica napus* stem. Secondly, the treatment time was different: Gellatly at al. [22] induced *L. maculans* cultures for 22 h while we harvested mycelium just 5 h after the induction to prevent secondary effects. It will be interesting to compare genes identified as induced by the treatment in these two experiments.

The group of Breyne et al. [7] recently published an attempt to further develop cDNA-AFLP technique into a quantitative genome-wide transcript profiling system. The major drawback of their protocol as compared to DDRT-PCR is a high fraction of transcripts which escape analysis because of the lack of suitable restriction sites (40% in the work by Breyne at al. [7]). This limitation casts doubts on the suitability of the system for genome-wide expression analysis. Furthermore, the detection and quantification of cDNA fragments was done by autoradiography of radioactively labeled DNA rather than on a DNA sequencer. The noise inherent to autoradiography is likely to increase statistical errors in intensity values assigned to bands. A crucial difference as compared to our approach concerns the normalization of band intensities. The authors calculated their normalization factor from the intensities of bands found to be expressed constantly during a time-course experiment while we automatized this process by calculating intensity ratios for all bands, sorting the bands by ratios and using bands from the 25% – 75% quartile for the calculation of a normalization factor. Our procedure is suitable for automatization and does not require data from time-course experiments, which are not available in the majority of DDRT-PCR applications.

## Conclusion

Our protocol extends the application of cDNA Differential Display (DDRT-PCR) from a gene discovery tool to quantitative transcript profiling. The system is based on a combination of widely accessible hardware (DNA sequencer and fluorescence scanner) and software (DNA fingerprinting analysis software and a spreadsheet program). Transcript profiles generated by quantitative DDRT-PCR according to our protocol consist of sets of numerical values (primer pair/fragment size vs. induction factor) which can be stored in a spreadsheet and easily compared with each other and extended by new experiments. DDRT-PCR thus offers an open end alternative for microarray-based transcriptomics, which is particularly attractive in research on non-model organisms where microarray analysis is not available or economical.

## Methods

### Cultivation of oilseed rape and phytoalexin elicitation

Seeds of the summer oilseed rape cultivar Lichosmos (Deutsche Saatveredelung, Lippstadt, Germany) were germinated in containers and transplanted after 7 days into 13 cm² pots. A mixture of steamed compost, sand and peat (3:1:1) was used as substrate. Plants were grown under semi-controlled conditions in the greenhouse (16 h photoperiod, 23°C). Leaves of the plants were removed at growth stage BBCH 61 (beginning of flowering) after about 9–10 weeks of cultivation. Stems were directly treated with a solution of 10 mM $CuCl_2$ until run-off point using a commercial garden sprayer. Plants were then transferred into a folia tunnel to keep high humidity. The treatment was repeated after 24 h. Finally, plants were harvested 48 h after the last treatment, cut into segments of about 5 cm and stored at -20°C.

### Extraction of plant secondary metabolites

Hundred grams of *Brassica napus* stems were homogenised with 200 ml of 0.1 M sodium phosphate buffer pH 7.2 in a vortex. The homogenate was filtered through paper filter and the cleared homogenate was extracted twice with an equal volume of ethyl acetate. Ethyl acetate fractions were combined and extracted with 1 volume of hexane. Hexane was removed on a rotary evaporator and the residue suspended in 30 ml of sterile Czapek-Dox medium (Oxoid Ltd, Hampshire, England) and stored at -80°C.

### Fungal cultures

Cultures of *Leptosphaeria maculans* T12aD34 [15] were started by inoculating $1 \times 10^7$ pycnidiospores into 35 ml of Czapek-Dox liquid medium (Oxoid Ltd, Hampshire, England) supplemented with 0.2% yeast extract (Difco Laboratories, Detroit, USA) in 300 ml Erlenmeyer flasks and incubated at 20°C in the dark without shaking.

### Treatment of fungal cultures and RNA isolation

10 ml of plant crude extract were added to 35 ml of 7-days old *Leptosphaeria maculans* cultures. 10 ml of sterile Czapek-Dox medium was added to control cultures. The incubation was continued for 5 h, after which the mycelium was separated from medium by filtration and frozen in liquid nitrogen. Total RNA was prepared using a guanidine thiocyanate/phenol method [16] modified as follows. Approximately 100 mg of mycelium was ground in a mortar under liquid nitrogen, transferred to a microcentrifuge tube containing 1 ml of GTC-buffer [16], vortexed for 10 s and incubated for 5–10 min at ambient temperature. After centrifugation for 5 min at $16,000 \times g$, the supernatant was transferred to a centrifuge tube containing 0.7 ml of phenol/chloroform/isoamylalcohol mixture (25:24:1), vortexed for 15 s and incubated on ice for 10 min. The emulsion was centrifuged for 10 min at $16,000 \times g$ at 4°C. The aqueous phase was transferred to a new tube and mixed with 0.2 vol of 1 M acetic acid and 0.7 vol of 100% EtOH. The suspension was incubated at -20°C for 30 min. Precipitated RNA was centrifuged for 10 min at $16,000 \times g$, 4°C. The pellet was washed with 70% ethanol twice, dried and homogenized with 0.7 ml of GTC buffer without mercaptoethanol. Undissolved residue was removed by centrifugation and RNA was precipitated with 0.14 ml of 3 M sodium acetate, pH 5.4 and 0.7 ml ethanol at -20°C for 30 min. The RNA pellet was collected by centrifugation, washed twice with 70% ethanol, dried and dissolved in 30 μl of DEPC-treated water.

RNA samples were treated with RNase-free DNase (Fermentas, St. Leon-Rot, Germany) to remove the residual DNA contamination as recommended by the manufacturer. The integrity and quantity of RNA was assessed on a 1%denaturing agarose gel [17].

### First strand cDNA synthesis

DNase treated total RNA was used for three cDNA synthesis reactions primed with Cy5-labeled anchored oligo-dT ($dT_{17}M$; M = dA, dG or dC). cDNA synthesis mixture was prepared by combining 5.5 μl DEPC-treated water, 4 μl of 5x reverse transcriptase buffer (Fermentas, St. Leon-Rot, Germany), 2 μl of dNTP (10 mM), 5 μl of DNase treated RNA (about 2 μg) and 2.5 μl of 40 μM Cy5-labeled anchored poly(dT)-primers. After incubation at 65°C for 5 min followed by 10 min at 37°C 200 U of reverse transcriptase (Maloney murine leukemia virus (MMLV)) (Fermentas, St. Leon-Rot, Germany) were added and the incubation was continued for 50 min. The reaction was arrested at 72°C for 5 min. The products were diluted by adding 80 µl of distilled water to each reaction.

### Polymerase chain reaction and gel electrophoresis

Second strand synthesis and PCR amplification were peformed in a total volume of 20 µl. The reaction mixture contained 20 µM dNTPs, 1 U of Taq Polymerase (GeneCraft, Muenster, Germany), 4 µM of arbitrary decamer deoxynucleotides (Operon, Cologne, Germany; for primer sequences see supplementary material), 0.5 µM Cy5-labeled anchored primer $(dT_{17}M)$ and 1 µl of diluted cDNA solution. PCR reactions with cDNA from the treated culture and control and each of 228 primer combinations (76 random primers vs. 3 anchored poly(dT) primers) were performed according to a published protocol [12] using an annealing temperature of 40°C and 26 cycles.

The whole PCR reaction was reduced to a suitable volume by incubating a PCR plate without a lid in a thermocylcer at 50°C for 1 h, mixed with 5 µl of loading solution (98% formamide, 10 mM EDTA, 0.025% bromophenol blue, 0.025% xylene cyanol FF, pH 8.0) [18] and denatured at 90°C for 3 min. A Cy5-labeled size standard from 50 to 500 nt (ALFexpress TM Sizer, Amersham Biosciences, Piscataway, USA) was used as size standard. Electrophoresis was performed on an ALFExpress II sequencer (Amersham Pharmacia Biotech AB, Uppsala, Sweden) with a 25 cm × 0.5 mm gel cassette and 6% polyacrylamide gel (Reprogel TM Long Read, Amersham Pharmacia Biotech AB, Uppsala, Sweden) at 25 W for 700 min at 55°C.

### Data processing

Detector signals from an ALFExpress II sequencer were converted into 16-bit TIFF images using ALF Sequence Analyzer 2.11 (Amersham Pharmacia Biotech AB, Uppsala, Sweden). Each chromatogram was exported using a low compression level (resampling factor of 2 was used for digital processing). Pseudogel images for digital processing were cropped to a length of 10,000 pixels by removing the lower part of the gel using Adobe Photoshop 6.0 (Adobe Systems Incorporated, San Jose, USA) and imported into the gel analysis software GelCompar 3.5 (Applied Maths, Kortrijk, Belgium). 10,000 pixel is the maximum gel length which can be processed by GelCompar. Gel image parts corresponding to fragments shorter than 50 bp were removed and lanes with artefacts were identified by manual inspection and discarded. A Cy5-labeled size standard (50-bp ladder) was defined as a standard marker lane on the first gel and the following gels were aligned to this marker.

Band finding was performed automatically using a minimum profiling value of 0.5%, allowing only for bands with an elevation of at least 0.5% with respect to the surrounding background. Minimum area and shoulder sensitivity were set to zero. Further editing was performed manually for those band positions in which a band was found only in one of the two corresponding lanes (treatment/control). Bands were added or discarded according to the following criteria: when the densitometric curve revealed an elevation at the position in which a band was found in the corresponding lane, a new band was introduced and verified by checking its shape fitted to the densitometric profile. In case no elevation of the densitometry profile was found in a position in which a band was found in the corresponding lane and the corresponding band was weak (elevation barely over 0.5%), the band was discarded. Cases of strong bands in one lane corresponding to no band in the same position in the corresponding lane were very rare. Band matching was performed automatically for each two lanes corresponding to the same primer combination with position tolerance set to 1%. It was rarely necessary to adjust the matching manually, for example when several adjacent bands largely overlapped or when the algorithm failed to find obvious matches because of a slight retardation of lanes at the margin of the gel. Absolute peak area and calculated fragment size for each band were exported to Microsoft Excel (Microsoft, Redmond, USA).

Normalization of densitometry values is crucial both for comparisons among profiles generated in the same experiment and among different experiments. Because the intensities of control and treatment lanes might differ due to differences in the efficiency of RNA extraction and RT-PCR and in the volumes loaded on the sequencer, band intensities (measured as peak areas) have to be adjusted by a normalization factor. This factor can be obtained as the mean of ratios of intensities of corresponding bands (treatment/control) for bands originating from transcripts that are unchanged by the treatment. A technical problem arises from the fact that a rigorous distinction between bands which are affected by the treatment and those which are not affected is only possible after the normalization, yet it is necessary to identify at least some unaffected bands for the calculation of the normalization factor. We solved the problem by assuming that only a minority of bands in each lane is affected by the treatment.

To identify bands which were not affected by the treatment for the estimation of the normalization factor, we calculated "uncorrected induction factors" (UIF) for each pair of bands in corresponding lanes as the ratio of areas of peaks corresponding to matched bands. In the next step UIFs for all bands in a lane were sorted by values.

**Table 2: Reproducibility of normalized band intensities**

| Intensity ratio | Proportion |
|---|---|
| 0.33 – 3.0 | 97.0% |
| 0.40 – 2.5 | 88.9% |
| 0.50 – 2.0 | 80.7% |

Peak areas of corresponding cDNA fragments derived from two independent control experiments generating 423 fragment pairs were normalized and compared to each other. Bands were assigned to intensity ratio classes defined in the left column.

Assuming that fewer than 25% of bands in a lane were induced and fewer than 25% suppressed by the treatment, the mean for all UIFs for bands between the first and third quartile was calculated. This value was used as a normalization factor to correct all absolute peak areas. Induction factors (IF) were now calculated as the ratio of normalized peak areas for matched bands in the treated sample and the control. Unaffected bands possessed induction factors close to 1.0, induced bands had IF > 1.0, suppressed bands had IF < 1.0.

*Statistics*

Data were analysed using Microsoft Excel 7 (Microsoft, Redmond, USA) and graphs drafted with SigmaPlot 5.0 (SPSS, Chicago, USA). For results presented as Box-and-Whisker plots, the boxes include 50% of the ranked data, the whiskers show the 10th and 90th percentile, the outlier points mark the outliers defined as values above and below the 10th and 90th percentile. Means are expressed as arithmetic mean ± S.D.

**Abbreviations**

cDNA – DNA complementary to RNA, generated by reverse transcription of mRNA; cDNA-AFLP – application of DNA-fingerprinting method Amplified Fragment Length Polymorphism to cDNA; Cy5 – fluorophore used for DNA labeling (dimeric unsymmetrical cyanine dye); DDRT-PCR – differential Display of cDNA, transcription profiling method based on the amplification of cDNA by a combination of a random primer and an anchored poly(dT)-primer; DEPC – diethylbiscarbonate; Dnase – desoxyribonuclase; dNTP – desoxyribonucleotide triphosphates; EDTA – ethylenediamminetetraacetic acid; GTC – guanidinium thiocyanate; Rnase – ribonuclease; IF – corrected induction factor; S.D. – standard deviation; UIF – uncorrected induction factor

**Authors' contributions**

BV performed all experiments except the induction of Brassica by copper. UH contributed to the design of algorithms for data processing, processed data and participated in drafting the manuscript. BK performed the induction of Brassica by copper and participated in data analysis and drafting the manuscript. PK conceived the idea, designed and coordinated the study, guided data processing and wrote major parts of the manuscript.

**Additional material**

**Additional File 1**

*Sequences of arbitrary primers* Excel-file containing the sequences of 80 primers used in combination with anchored poly(dT)-primers in DDRT-PCR reactions. 19 KB; http://www.biomedcentral.com/imedia/6962194924713244/sup1.xls
Click here for file
[http://www.biomedcentral.com/content/supplementary/1471-2164-6-51-S1.xls]

**References**

1. Brenner S, Johnson M, Bridgham J, Golda G, Lloyd DH, Johnson D, Luo S, McCurdy S, Foy M, Ewan M, Roth R, George D, Eletr S, Albrecht G, Vermaas E, Williams SR, Moon K, Burcham T, Pallas M, DuBridge RB, Kirchen J, Fearon K, Mao J, Corcoran K: **Gene expression analysis by massively parallel signature sequencing (MPSS) on microbead arrays.** *Nat Biotechnol* 2000, **18:**630-634.
2. Shimkets RA, Lowe DG, Tai JT, Sehl P, Jin H, Yang R, Predki PF, Rothberg BE, Murtha MT, Roth ME, Shenoy SG, Windemuth A, Simpson JW, Daley MP, Gold SA, McKenna MP, Hillan K, Went GT, Rothberg JM: **Gene expression analysis by transcript profiling coupled to a gene database query.** *Nat Biotechnol* 1999, **17:**798-803.
3. Velculescu VE, Zhang L, Vogelstein B, Kinzler KW: **Serial analysis of gene expression.** *Science* 1995, **270:**484-487.
4. Liang P, Pardee A: **Differential display of eukaryotic messenger RNA by means of the polymerase chain reaction.** *Science* 1992, **257:**967-971.
5. Bachem CWB, van der Hoeven RS, de Bruijn SM, Vreugdenhil D, Zabeau M, Visser RGF: **Visualization of differential gene expression using a novel method of RNA ?ngerprinting based on AFLP: analysis of gene expression during potato tuber development.** *Plant J* 1996, **9:**745-753.
6. Liang P: **A decade of differential display.** *BioTechniques* 2002, **33:**338-346.
7. Breyne P, Dreesen R, Cannoot B, Rombaut D, Vandepoele K, Rombauts S, Vanderhaeghen R, Inze D, Zabeau M: **Quantitative cDNA-AFLP analysis for genome-wide expression studies.** *Mol Genet Genomics* 2003, **269:**173-179.
8. Stein S, Liang P: **Differential display analysis of gene expression in mammals: a p53 story.** *Cell Mol Life Sci* 2002, **59:**1274-1279.
9. Jones SW, Cai D, Weislow OS, Esmaeli-Azad B: **Generation of multiple mRNA fingerprints using fluorescence-based differential display and an automated DNA sequencer.** *BioTechniques* 1997, **22:**536-543.
10. Luehrsen KR, Marr LL, van der Knaap E, Cumberledge S: **Analysis of differential display RT-PCR products using fluorescent primers and GENESCAN software.** *BioTechniques* 1997, **22:**168-174.
11. Cho Y, Meade J, Walden J, Guo Z, Liang P: **Multicolor fluorescent differential display.** *BioTechniques* 2001, **30:**562-572.
12. Ito T, Kito K, Adati N, Mitsui Y, Hagiwara H, Sakaki Y: **Fluorescent differential display: arbitrarily primed RT-PCR fingerprinting on an automated DNA sequencer.** *FEBS Lett* 1994, **351:**231-236.
13. Qin L, Prins P, Jones JT, Popeijus H, Smant G, Bakker J, Helder J: **GenEST, a powerful bidirectional link between cDNA sequence data and gene expression profiles generated by cDNA-AFLP.** *Nucleic Acids Res* 2001, **29:**1616-1622.
14. Lyne R, Burns G, Mata J, Penkett CJ, Rustici G, Chen D, Langford C, Vetrie1 D, Bähler J: **Whole-genome microarrays of fission yeast: characteristics, accuracy, reproducibility, and processing of array data.** *BMC Genomics* 2003, **4:**27.

15. Kuswinati T, Koopmann B, Hoppe HH: **Virulence pattern of aggressive isolates of** *Leptosphaeria maculans* **on an extended set of** *Brassica* **differentials.** *J Plant Disease Protect* 1999, **106:**12-20.

16. Chomczynski P, Sacchi N: **Single step method of RNA isolation by acidguanidinium-thiocyanate-phenol-chloroform extraction.** *Anal Biochem* 1987, **162:**156-159.

17. Sambrook J, Fritsch EF, Maniatis T: *Molecular Cloning: A Laboratory Manual* 2nd edition. Cold Spring Harbor, NY: Cold Spring Harbor Laboratory Press; 1989.

18. Reineke A, Karlovsky P: **Simplified AFLP protocol: replacement of primer labeling by the incorporation of α-labeled nucleotides during PCR.** *BioTechniques* 2000, **28:**622-623.

19. Liang P, Pardee A: *Differential display methods and protocols* Totowa, NJ: Humana Press Inc; 1997.

20. Bauer D, Muller H, Reich J, Riedel H, Ahrenkiel V, Warthoe P, Strauss M: **Identification of differentially expressed mRNA species by an improved display technique (DDRT-PCR).** *Nucleic Acids Res* 1993, **21:**4272-4280.

21. Aittokallio T, Ojala P, Nevalainen TJ, Nevalainen O: **Automated detection of differentially expressed fragments in mRNA differential display.** *Electrophoresis* 2001, **22:**1935-1945.

22. Gellatly KS, Ash GJ, Taylor JL: **Development of a method for mRNA differential display in filamentous fungi: comparison of mRNA differential display reverse transcription polymerase chain reaction and cDNA amplified fragment length polymorphism in** *Leptosphaeria maculans.* *Can J Microbiol* 2001, **10:**955-960.

23. George K, Zhao X, Gallahan D, Shirkey A, Zareh A, Esmaeli-Azad B: **Capillary electrophoresis methodology for identification of cancer related gene expression patterns of fluorescent differential display polymerase chain reaction.** *J Chromatogr B Biomed Sci Appl* 1997, **695:**93-102.

24. Aittokallio T, Pahikkala T, Ojala P, Nevalainen TJ, Nevalainen O: **Electrophoretic signal comparison applied to mRNA differential display analysis.** *Biotechniques* 2003, **34:**116-122.

25. Short JM, Fernandez JM, Sorge JA, Huse WD: **Lambda ZAP: a bacteriophage lambda expression vector with in vivo excision properties.** *Nucleic Acids Res* 1988, **16:**7583-7600.